\documentclass[12pt]{iopart}

\usepackage{mathpazo}
\usepackage{color,cancel,wrapfig,lscape,array}
\usepackage{textcomp}
\usepackage{times}
\usepackage{ragged2e}
\usepackage{mathptmx}
\usepackage{hhline}
\usepackage{enumerate}
\usepackage{bbm}
\usepackage{units}
\usepackage{nicefrac}
\usepackage{graphicx}
\usepackage[dvips]{rotating}
\usepackage{color}
\usepackage{subfigure}
\definecolor{violet}{cmyk}{0.94,0.86, 0.04,0.08}

\usepackage{iopams}
\expandafter\let\csname equation*\endcsname\relax
\expandafter\let\csname endequation*\endcsname\relax
\usepackage{amsmath,amsfonts,amssymb,dsfont,amsxtra,amsopn,amstext,amsbsy}
\usepackage{amssymb,amsthm,amsthm,amsfonts}
\usepackage[T1]{fontenc}
\usepackage[latin1]{inputenc}

\usepackage{xr}

\newcommand{\ndg}{{\phantom{\dagger}}}
\newcommand{\dg}{\dagger}
\newcommand{\ew}[1]{\pmb{\big\langle} #1 \pmb{\big\rangle}}

\newcommand{\ket}[1]{| #1 \rangle}
\newcommand{\bra}[1]{\langle #1 |}

\newcommand{\ketbra}[2]{\left|#1\right\rangle\hspace{-1.1mm}\left\langle #2 \right|}

\newcommand{\sig}[2]{\sigma_{#1 #2}}

\begin{document}

\title{Correlations of cascaded photons: An analytical study 
in the two-photon Mollow regime}

\author{Alexander Carmele, Samir Bounouar, Max Strau\ss, Stephan Reitzenstein, 
and Andreas Knorr}
\address{Nichtlineare Optik und Quantenelektronik, Institut f\"ur Theoretische 
Physik \\ 
Optoelektronik und Quantenbauelemente, Institut f\"ur Festk\"orperphysik \\ 
Technische Universit\"at Berlin, Hardenbergstr. 36, 10623 Berlin}
\ead{alex@itp.tu-berlin.de}

\begin{abstract} 
Mollow physics in the two-photon regime shows interesting features
such as path-controlled time-reordering of photon pairs without the need
to delay them. 
Here, we calculate analytically the two-photon correlations $ g^{(2)}(\tau)$,
essential to discuss and study such phenomena in the resonant-driven 
dressed-state regime.
It is shown that there exists upper and lower bounds of the $ g^{(2)}(\tau)-$
function for certain spectrally-selected photon pairs. 
Recent reported experiments agree with the presented theory and 
thereby it is shown that the resonant-driven four-level system is an 
interesting source for steerable quantum light in quantum cascade setups.
We furthermore discuss the unlikeliness to observe antibunching
for the delay time $ \tau=0 $ in the exciton-biexciton 
correlation functions in such experiments, since antibunching stems from a
coherent and in-phase superposition of different photon emission events.
Due to the occuring laser photon scattering, this coherent
superposition state is easily disturbed and leads to correlation
functions of $ g^{(2)}(0)=1$.
\end{abstract}

\section{Introduction}\label{sec:introduction}
Two-photon states are of particular interest to study
non-linear phenomena in quantum optics and lie at the 
heart of the pursuit to exploit nonlocal properties 
in quantum mechanics itself 
\cite{mandel1999quantum,scully,carmichael2009statistical,gardiner2015quantum}. 
Two-photon states generated in an atomic cascade
allowed the first experiments to test Bell's inequality
\cite{freedman1972experimental,
clauser1978bell,aspect1981experimental,aspect1982experimental} 
and have paved the way to significant-loophole free tests 
in other material platforms recently 
\cite{giustina2015significant,hensen2015loophole}.
Nowadays semiconductor based biexciton-cascade processes generate 
on demand polarized-entangled photon pairs to a very high degree
of purity due to recent progresses in nanotechnology fabrication
\cite{stevenson2006semiconductor,salter2010entangled,muller2014demand,
winik2017demand}.
They also gave rise to test the indistinguishability of 
photons created in the spontaneous parametric downconversion
via Hong-Ou-Mandel type of 
experiments \cite{hong1987measurement,ou1988violation,mandel1999quantum}, 
and semiconductor growth techniques allow now high extraction efficiencies
and close to perfect visibilities 
\cite{santori2002indistinguishable,thoma2016exploring,
ding2016demand,muller2017quantum}.
Furthermore, two-photon signatures reveal the true
quantum character of cavity quantum electrodynamics via
anharmonic, second-rung features that scale non-linearly with the 
photon number. 
Experimentally, higher-order rungs have been extracted in
superconducting circuit platforms \cite{fink2008climbing,fink2010quantum}, 
atom cavity-QED systems \cite{PhysRevLett.76.1800,hamsen2017two} and 
also in semiconductor nano-emitters clear signatures have been observed
\cite{kasprzak2010up,hopfmann2017transition}. 
Thus, two-photon sources have a wide application domain 
and allow to test a broad range of quantum behavior. 
It is therefore highly feasible to understand two-photon
generation processes based on incoherent and coherent driving
on an analytical and fundamental level.
Recently, a proposal has drawn interest, which allows one to 
coherently generate photon bundles in leap-frog processes
\cite{del2012theory,munoz2014emitters,munoz2015enhanced}.
The idea behind is to drive the emitter in resonance with 
the N-photon transition but off-resonant with a single 
photon coherence \cite{ota2011spontaneous,schumacher2012cavity}.
This way, single-photon processes are suppressed which tend
to degrade two- or N-photon coherences.
Furthermore, the polarization degree of freedom can be used 
to probe only emission processes without scattered photons
from the driving laser field. 
This idea has been experimentally realized recently 
with a semiconductor quantum dot 
\cite{hargart2016cavity,ardelt2016optical,bounouar2016time} .
Here, the biexciton is coherently driven in a two-photon process
and via only one of the two possible polarization transitions.
In the strong driving regime, dressed state become visible and 
allow for frequency-selected photon-photon correlation measurements
\cite{bounouar2016time}.
This way, path-controlled time-reordering of photon pairs 
can be steered.
As a measure whether there is and there is no time-order, we
use the correlation function and investigate whether the function
is symmetric under sign exchange $g^{(2)}(\tau)=g^{(2)}(-\tau)$ 
or not.
If the correlation function is symmetric, the time-order of the 
emission sequence is erased. 
Such a vanishing time-reordering is only possible, when the emitter 
is driven into a superposition state with long coherence times.
Here, we give the theoretical background for the experiment 
\cite{bounouar2016time} and calculate the correlation functions 
analytically.
We discuss symmetries and also limits for given measurement 
detection sequences and show, when and under which circumstances 
perfect time-reordering can be reached, or when anti-correlation is 
inevitable and bunching will always be the case.
We also reveal, which values of the correlation functions values are 
inaccessible for certain measurement sequences and provide therefore 
a toolbox 
to interpret these kind of multi-photon processes with the correlation 
function data at hand. 
The analysis below shows that the physics behind this experiment 
can still be described with two-level physics, but 
with a two-photon
substructure 
\cite{carmichael2009statistical,PhysRevLett.58.2539,schrama1992intensity,
roy2012polaron, PhysRevB.78.153309}.
The article is organized as follows:
After this introduction, cf.~\ref{sec:introduction},
we continue in Sec.~\ref{sec:model} with presenting 
the model and discuss the Hamiltonian of the system.
The chosen excitation frequency allows to adiabatically eliminate the 
the exciton levels of the four-level system up to moderate driving 
strengths.
Without the excitonic degrees of freedom, the effective model is 
isomorph to a two-level system, which can be solved in various ways.
We solve it via a Laplace transformation in contrast to most
diagonalization approaches common in literature, e.g. \cite{scully}.
These solutions are interpreted with respect to their two-photon 
substructure and used to calculate the photon-photon correlation 
function in the bare state basis in Sec.~\ref{sec:bare_state_g2}.
The results will be a very asymmetric correlation function around 
$ \tau=0 $ with antibunching to one side and bunching to the other
side.
However, in the dressed state basis, derived in 
Sec.~\ref{sec:dressed_state_basis}, certain contributions show a 
symmetric behavior.
Sec.~\ref{sec:dressed_state_g2} is dedicated to the calculation
of the correlation in the dressed-state basis. 
Those are the main result of the paper and explain the 
experimental data of 
Ref.~\cite{bounouar2016time} in case of resonant two-photon excitation.
To complete the discussion, intensity dependent experimental data
are qualitatively compared to the theory in Sec.~\ref{sec:experiment}.
The symmetrization of the correlation is clearly seen, now supported
additionally via the analytical calculations, before we conclude in 
Sec.~\ref{sec:conclusion}.
%

\begin{figure}[t!]
\centering
\includegraphics[width=16cm]{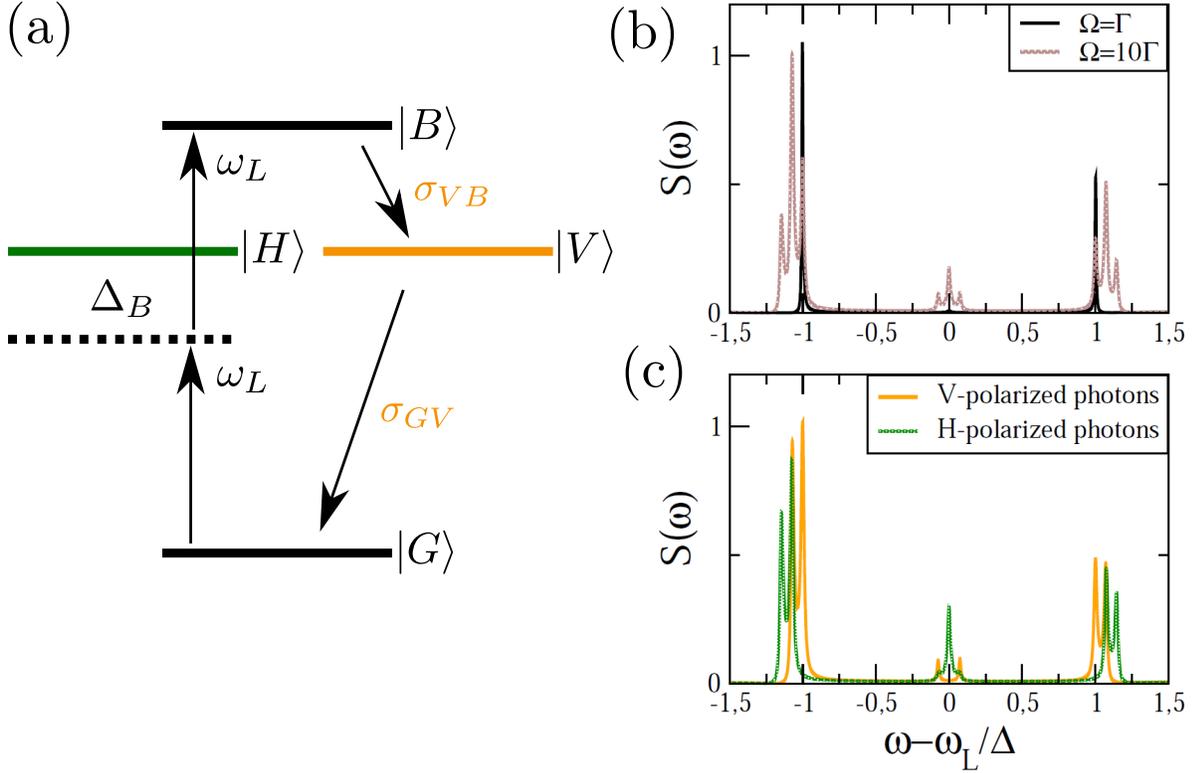}
\caption{(a) The energy schema of the setup. The biexciton state
is driven via a continuous two-photon, $ H-$polarized excitation, while the 
exciton transitions are far detuned. (b) The power spectrum $S(\omega)$
of the full emission, including $H-$ and $ V-$polarized photons for 
two excitation strengths. In the strong excitation limit (grey),  
triplets appear. (c) Differentiating the spectrum in the strong excitation
limit between $H-$ (green,shaded) and $ V-$ emission (orange, solid) events.
Due to the ignored fine structure splitting $ \Delta_\text{FSS}=0$ two 
peaks overlap.}
\label{fig:schema}
\end{figure}

\section{Model}\label{sec:model}
The quantum dot is modelled as a four-level system consisting of a 
ground state $\ket{G}$, two excitonic states $ \ket{H},\ket{V}$ 
and a biexcitonic state $ \ket{B} $.
The ground state energy $ \omega_G=0 $ is set to zero, 
cf.~Fig.~\ref{fig:schema}.
The two excitonic states may differ in their energies 
$ \omega_H-\omega_V=\Delta_\text{FSS}$ due to the fine structure
splitting.
This splitting is of great importance in experiments probing 
photon polarization entanglement generated in a biexciton cascade.
Here, we focus on a driven experiment, where photon correlations 
in the Mollow regime are studied and the fine structure splitting 
is of minor importance. 
So, we set it, out of convenience, to zero for the following 
discussion: $ \Delta_\text{FSS}=0$, i.e. $ \omega_H=\omega_V=\omega_X$.
In contrast, the biexciton shift is of particular interest, as only 
due to this shift it is possible to drive the biexciton population 
without driving necessarily at the same time the excitonic transitions.
The biexciton shift ($\Delta_B$) can be attractive or repulsive and is 
defined as 
the difference between the sum of the exciton energies and the bare 
biexciton resonance, rendering the biexciton energy 
$\omega_B=\omega_H+\omega_V+2\Delta_B=2(\omega_X+\Delta_B)$.
Here, the external laser field drives only the horizontal 
polarization with frequency $ \omega_L $ and 
amplitude $ \Omega_H $, the full Hamiltonian reads therefore 
$(\hbar=1)$ and $ \sig{i}{j}:=\ketbra{i}{j}$:
\begin{align}
H =& 
\omega_X 
\left(\sig{H}{H}+ \sig{V}{V}\right)
+
2\left(\omega_X+\Delta_B\right)\sig{B}{B} 
+ \Omega_H \cos(\omega_Lt) 
( \sig{G}{H} + \sig{H}{G} + \sig{B}{H} + \sig{H}{B} ). 
\end{align}
To allow for an adiabatic elimination of the excitonic states in case 
of biexciton shift in the regime $ \Delta_B \gg \Omega_H $, the 
laser frequency is chosen to be in a two-photon resonance with the 
biexciton frequency, namely we set 
$ \omega_L = \omega_H/2+\omega_V/2+\Delta_B$.
This leads to the following Hamiltonian, assuming $ \omega_H=\omega_V $,
and after the rotating-wave approximation and transforming into the 
rotating frame of the laser frequency:
\begin{align}
H_R =& 
\Delta \left( \sig{H}{H} + \sig{V}{V}  \right)
+ 
\Omega_L 
(\sig{G}{H} + \sig{H}{G} + \sig{B}{H} + \sig{H}{B}), 
\end{align}
with $ \Delta=-\Delta_B$ and $ \Omega_L=\Omega_H/2$, 
cf.~Fig.~\ref{fig:schema}(a).
We assume a radiative decay of the electronic system via photon 
emission into a Markovian continuum to yield the following master 
equation:
\begin{align}
\dot \rho =& -i \left[H_R,\rho\right] 
+ \Gamma_X
\left( 
\mathcal{D}[\sig{G}{H}]
+
\mathcal{D}[\sig{G}{V}]
+
\mathcal{D}[\sig{B}{H}]
+ 
\mathcal{D}[\sig{B}{V}]
\right)\rho, \label{eq:full_meq}
\end{align}
after assuming the biexciton decay to be double as fast as the exciton 
decay and using the standard Lindblad 
form $\mathcal{D}[J]\rho=2 J\rho J^\dg-\lbrace J^\dg J,\rho\rbrace$.
Pure dephasing contributions are neglected on this 
timescale~\cite{richter2009two} and are set in the master equation, 
out of convenience, to zero. 
The following analytical calculations can be done with a finite pure
dephasing,rendering them however more lengthy. 
Since we focus here on the intensity dependent photon-photon correlation 
in the low temperature limit $ T \approx 4K $, where the dissipation 
in the system is mainly governed by the radiative decay, the pure 
dephasing is assumed to be of minor importance for the studied 
correlation function, which is supported also by the intensity-dependent 
experimental data, in Sec.~\ref{sec:experiment}. 
In Fig.~\ref{fig:schema}(b) and (c), the power spectrum of the system
is plotted: 
$ S(\omega)=
\lim\limits_{t \to \infty}
\int_0^\infty \ew{c^\dg(t)c(t+\tau)}e^{-i\omega\tau}$ with $ 
c(t)=\sig{i}{B}(t)+\sig{G}{i}(t)$ for $ i=H,V$.
For weak excitation, only two emission peaks from the bare exciton and 
bare biexciton photons are visible, cf.~Fig.~\ref{fig:schema}(b, black line),
since the Mollow splitting is not large enough in comparison to the 
radiative linewidth.
If the driving strength increases, triplets appear around the bare 
resonances (gray line) and around the laser frequency.
To distinguish the contributions, a polarization filter can be applied.
The $H-$polarized photons show a very different dependence on the 
driving strength than the $V-$polarized photons. 
In contrast to the latter, both peaks of the $H-$polarized photons shift 
with the driving strength (green, shaded).
However, only one peak of the $V-$polarized photons shift (orange solid),
and one peak stays independently on the driving strength on the bare 
resonance.
In the following, the time dynamics of the density matrix is analytically 
solved. 
Using the master equation, the following set of differential equations of
motion can be derived with $ \rho_{ij}=\bra{i}\rho\ket{j}$:
\begin{align}
\dot \rho_{GG} 
=&
2 \Gamma_X (\rho_{HH}+\rho_{VV})
-i\Omega_L
\left( 
\rho_{HG}
-
\rho_{GH}
\right), \\
\dot \rho_{HG} 
=&
-
\left(
\Gamma_X 
+i\Delta
\right)
\rho_{HG}
-i\Omega_L
\left( 
\rho_{BG}
+
\rho_{GG}
-
\rho_{HH}
\right), \\
\dot \rho_{HH} 
=&
-2 \Gamma_X 
\left( \rho_{HH} - \rho_{BB} \right)
-i\Omega_L
\left( 
\rho_{GH}
-
\rho_{HG}
+
\rho_{BH}
-
\rho_{HB}
\right), \\
\dot \rho_{BG} 
=&
-
2\Gamma_X 
\rho_{BG}
-i\Omega_L
\left( 
\rho_{HG}
-
\rho_{BH}
\right), \\
\dot \rho_{BH} 
=&
-
\left(
3
\Gamma_X 
-i\Delta
\right)
\rho_{BH}
-i\Omega_L
\left( 
\rho_{HH}
-
\rho_{BB}
-
\rho_{BG}
\right), \\
\dot \rho_{BB} 
=&
-4 \Gamma_X \rho_{BB}
-i\Omega_L
\left( 
\rho_{HB}
-
\rho_{BH}
\right), \\
\dot \rho_{VV} 
=&
-2 \Gamma_X \left( \rho_{VV}-\rho_{BB}\right).  
\end{align}
Note, the $V$-exciton level is not driven by the external
driving field, thus $ \rho_{VG} $ and $ \rho_{VB} $ and its 
complex conjugates are no dynamical quantities.
Now, we choose a particular quantum dot, where the biexciton
shift is much larger than the radiative decay constants, which 
is typically the case $\Delta \gg \Gamma_X$.
Furthermore, we choose also a driving strength much weaker than 
the biexciton binding energy, therefore $\Delta \gg \Omega_L$.
Therefore, we can eliminate the transition to the 
$ H $ exciton, as the time scale is dominated by the 
detuning:
\begin{align}
\rho_{HG} 
=&
\frac{i\Omega_L}{\Gamma_X+i\Delta}
\left[
\rho_{HH}
-
\rho_{BG}
-
\rho_{GG} 
\right]
\approx 
\frac{\Omega_L}{\Delta}
\left[
\rho_{HH}
-
\rho_{BG}
-
\rho_{GG} 
\right] \\
\rho_{BH} 
=&
\frac{i\Omega_L}{3\Gamma_X-i\Delta}
\left[
\rho_{BB}
+
\rho_{BG}
-
\rho_{HH} 
\right]
\approx 
-
\frac{\Omega_L}{\Delta}
\left[
\rho_{BB}
+
\rho_{BG}
-
\rho_{HH} 
\right], 
\end{align}
as we can safely assume $ \Delta \gg 3\Gamma_X $, as
the biexciton energy is in the regime of meV, and the radiative 
decay constant typically in the domain of $\mu$eV.
If we apply these adiabatical solutions into the 
corresponding equations, and again, ignore the 
dispersive shifts $ \Omega_L^2/\Delta \ll \Gamma_X $, which 
limits the validity of the following equations to 
medium driving strength, we yield:
\begin{align}
\dot \rho_{BB} 
=&
-4\Gamma_X
\rho_{BB}
+
2
\text{Im}
\left[ 
\frac{\Omega_L^2}{\Delta}
\rho_{BG}
\right],  \\
\dot \rho_{GG} 
=&
2\Gamma_X
\left( 
\rho_{HH}
+ 
\rho_{VV}
\right)
-
2
\text{Im}
\left[ 
\frac{\Omega_L^2}{\Delta}
\rho_{BG}
\right],  \\
\dot \rho_{BG} 
=&
-2\Gamma_X
\rho_{BG}
-i
\frac{\Omega_L^2}{\Delta}
\left( 
\rho_{BB}
-
\rho_{GG}
\right).
\end{align}
Note, the excitonic densities are no longer a dynamical
quantity, they just follow the dissipative dynamics of the
biexciton state.
So, we can use $\rho_{HH}+\rho_{VV}=\Sigma_0-\rho_{BB}-\rho_{GG}$
with typically $\Sigma(t=0)=\Sigma_0=1$ and define $\Omega := 2 
\frac{\Omega_L^2}{\Delta}$
as the effective two-photon Rabi frequency and $ \Gamma:=2\Gamma_X$.
We have denoted the sum of the initial occupations $\Sigma_0$ to include
later consistently the solution for different initial conditions
in terms of the quantum regression theorem.
The solution is easier to calculate for the inversion, i.e. 
the population difference between biexciton and ground state
$ D(t) := \rho_{BB}(t) - \rho_{GG}(t)$,
and the linear independent imaginary part of the polarization
between both states
$B(t) := \rho_{BG}(t) - \rho_{GB}(t) $.
We yield a closed set of equations of motion:
\begin{align}
\dot D(t) 
=& 
-\Gamma \left[ D(t) + \Sigma_0 \right]- i\Omega B(t) \label{eq:dt} \\
\dot B(t) 
=& 
-\Gamma B(t) - i\Omega D(t) \label{eq:bt} .
\end{align}
Suprisingly, this set of equations is identical 
with a continuously driven two-level system, 
see \ref{app:two_level_compare}.
We conclude, that the two-photon driven biexciton system is 
in the adiabatic limit isomorph to the two-level dynamics.
However, the underlying structure of the four levels is still
present due to the conservation of angular momentum, as the 
biexciton cannot be driven directly from the ground state but
only via an intermediate excitonic state, i.e. $ \sigma_{GB}$
corresponds not to a electronic dipole operator.
We apply the Laplace transform and rederive the solution of the 
well-known Mollow problem, see \ref{app:laplace_solution}.
In the time domain, the inversion dynamics are given:
\begin{align}
D(t)
=& 
\left(
D_0
+\Sigma_0\Gamma\Gamma_n
\right)
e^{-\Gamma t} 
\cos(\Omega t)
-
\left(
iB_0
+\Sigma_0\Gamma\Omega_n
\right)
e^{-\Gamma t} \sin(\Omega t)
-\Sigma_0\Gamma\Gamma_n,
\end{align}
with $ \Gamma_n=\Gamma/(\Gamma^2+\Omega^2)$
and $ \Omega_n=\Omega/(\Gamma^2+\Omega^2)$.
This equation fulfills the initial condition $ D(t=0)=D_0 $.
The general equation of the polarisation dynamics reads:
\begin{align}
B(t) =& 
\left(
B_0 
-i\Sigma_0\Gamma\Omega_n
\right)
e^{-\Gamma t} \cos(\Omega t)
-i
\left(
\Sigma_0
\Gamma\Gamma_n
+
D_0 
\right)
e^{-\Gamma t} \sin(\Omega t)
+
i
\Sigma_0
\Gamma
\Omega_n,
\end{align}
with $ B(t=0)=B_0 $. 
Given these explicit solutions of the dynamics, all quantities can be
derived via integration.
For example, the biexciton dynamics reads:
\begin{align}
\rho_{BB}(t) 
=&
\rho_{BB}(0) 
e^{-2\Gamma t}
-iB_0 
\frac{\Omega}{2}
e^{-\Gamma t}
\left[
 \Gamma_n\cos(\Omega t)
+\Omega_n\sin(\Omega t)
-\Gamma_n
e^{-\Gamma t}
\right] 
\\
&
-
D_0
\frac{\Omega}{2}
e^{-\Gamma t}
\left[
 \Gamma_n\sin(\Omega t)
-\Omega_n\cos(\Omega t)
+\Omega_n e^{-\Gamma t}
\right] \\
&
+
\Sigma_0
\frac{\Omega}{2}
e^{-\Gamma t}
\left[
\Omega_n
\sinh(\Gamma t)
-\Gamma_n
\sin(\Omega t)
\right].
\end{align}
The detailed calculation is given 
in~\ref{app:explicit_general_solutions}.
So, with the biexciton dynamics at hand, we can derive the ground state
dynamics easily due to the definion of $ D(t) $:
\begin{align}
\rho_{GG}(t) 
=&
\rho_{BB}(t)
-
D_0
e^{-\Gamma t} \cos(\Omega t)
+iB_0
e^{-\Gamma t} \sin(\Omega t)
\\
&
+
\Sigma_0
\Gamma 
\left[
\Gamma_n
+
e^{-\Gamma t}
\
\left(
\Omega_n\sin(\Omega t)-\Gamma_n\cos(\Omega t)
\right)
\right].
\end{align}
For the measurements in the following, the $V-$exciton population
is also of importance. 
If the initial values of both excitonic states are equal, 
their dynamics agree in the adiabatic limit, as the differential 
equations in spite of the driving are effectively 
the same, so one can use $ \rho_{HH}(t)=\rho_{VV}(t)$ 
in case of $\rho_{HH}(0)=\rho_{VV}(0)$.
Then, one can conveniently use 
\begin{align}
\rho_{VV}(\tau)
=\frac{1}{2}\left(\Sigma_0-\rho_{GG}(\tau)-\rho_{BB}(\tau)\right)
=\frac{1}{2}\left(\Sigma_0+D(\tau)-2\rho_{BB}(\tau)\right), 
\end{align}
again typically $\Sigma=1$ but not if the conditional probability
in two-time-correlations is calculated.
However, if their initial values do not agree, one has to calculate 
the dynamics explicitly via
\begin{align}
\rho_{VV}(t) 
=&
\rho_{VV}(0) 
e^{-\Gamma t}
+
\Gamma 
e^{-\Gamma t}
\int_0^t
dt^\prime
e^{\Gamma t^\prime}
\
\rho_{BB}(t^\prime).
\end{align}
Finally, the polarisation reads:
\begin{align}
\rho_{BG}(t) 
=& 
\frac{1}{2}
B(t)
+
\frac{1}{2}
e^{-\Gamma t}
(
\rho_{BG}(0)
+
\rho_{GB}(0)
)
\end{align}
In the following, the exciton-biexciton photon correlations are calculated
explicitly to unravel the dependence on the two-photon driving strength 
$ \Omega=2\Omega_L^2/\Delta $ in comparison to the radiative decay 
constant $ \Gamma=2\Gamma_X $.
%

\section{Biexciton-Exciton Photon-Photon Correlation}\label{sec:bare_state_g2}
%
The measurement setup discriminates between the biexciton and exciton
photon through the different frequencies but with the same polarization,
here $V$, the undriven transitions.
The generation of photons is in the far field proportional to 
the corresponding transition operator $c^\dg_{ij}(t)=\sig{i}{j}(t)$ 
\cite{scully}.
The corresponding intensity-intensity correlation reads:
\begin{align}
g^{(2)}_{ijkl}(\tau) =& 
\lim\limits_{t \to \infty}
\frac{\ew{c^\dg_{ij}(t)c^\dg_{kl}(t+\tau)c^\ndg_{lk}(t+\tau)c^\ndg_{ji}(t)}}
{\ew{c^\dg_{ij}(t)c^\ndg_{ji}(t)}\ew{c^\dg_{kl}(t)c^\ndg_{lk}(t)} }.
\end{align}
In the density matrix picture, and with the corresponding transition
operators, the correlations function is given with
\begin{align}
\notag
& \lim\limits_{t \to \infty}
\ew{c^\dg_{ij}(t)c^\dg_{kl}(t+\tau)c^\ndg_{lk}(t+\tau)c^\ndg_{ji}(t)}
\equiv
\lim\limits_{t \to \infty}
\text{Tr}
\left[\rho(0)
\sig{i}{j}(t)\sig{k}{l}(t+\tau)\sig{l}{k}(t+\tau)\sig{j}{i}(t)
\right] 
\\ 
&=
\lim\limits_{t \to \infty}
\text{Tr}
\left[
\ketbra{j}{i}\rho(t)\ketbra{i}{j}\sig{k}{l}(\tau)\sig{l}{k}(\tau)
\right] 
=
\text{Tr}
\left[
\ketbra{j}{j}\rho_{ii}(\infty)\sig{k}{l}(\tau)\sig{l}{k}(\tau)
\right]= 
\notag \\ 
&
=
\rho_{ii}(\infty)
\text{Tr}
\left[
\rho^{jj}(\tau)
\ketbra{k}{l}
\ketbra{l}{k}
\right] 
=
\rho_{ii}(\infty)
\bra{k}\rho^{jj}(\tau)\ket{k}
=
\rho_{ii}(\infty)\rho^{jj}_{kk}(\tau), 
\end{align}
where $ \rho^{jj}=\ketbra{j}{j}$ is the conditional density matrix after the 
first measurement finding the state $ \ket{i}$. 
%
\begin{figure}[t!]
\centering
\includegraphics[width=16cm]{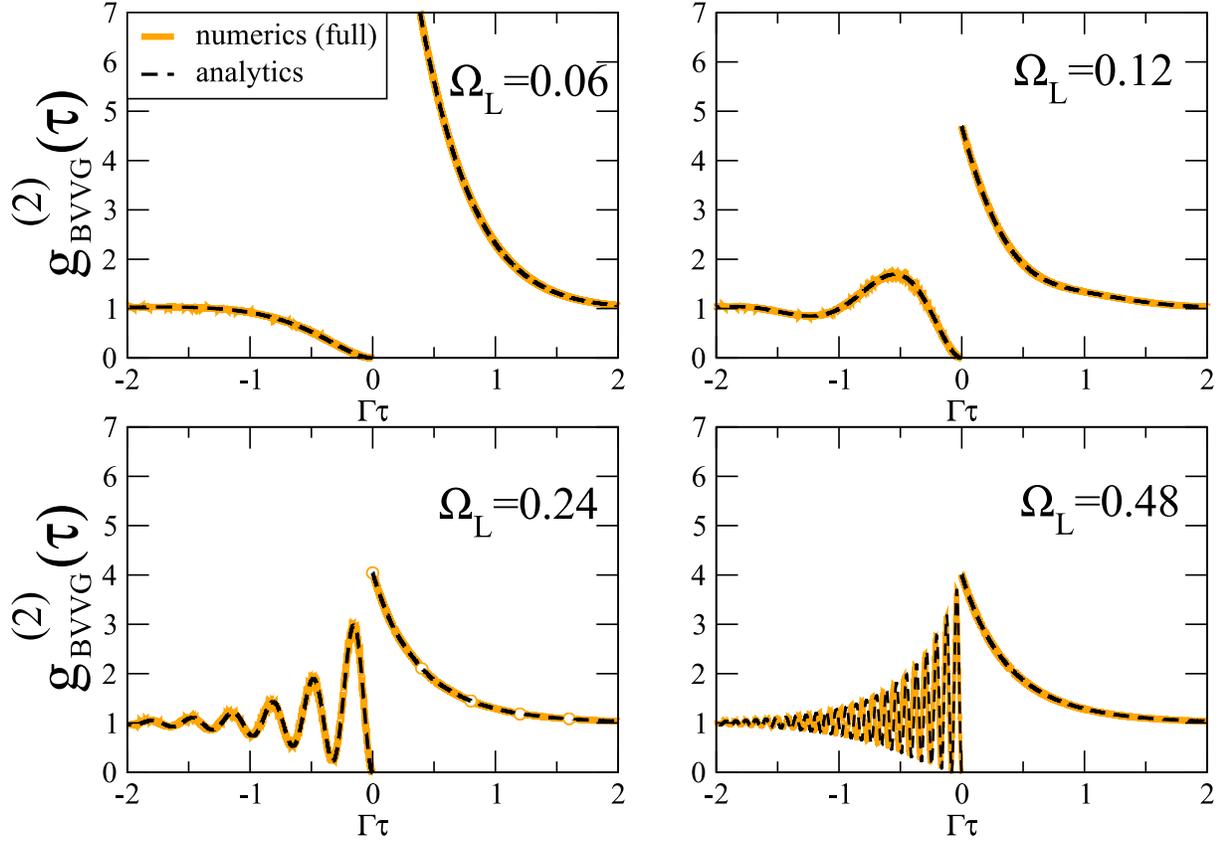}
\caption{Photon-photon correlations between exciton and biexciton 
photons for different driving strengths $ \Omega_L[\text{ps}^{-1}]$ with 
$ \Gamma=0.002\text{ps}^{-1}$ and $ \Delta=3.0\text{ps}^{-1}$. The 
numerical solution is derived from the complete master equation. The 
analytical solutions are given in the text in the adiabatic limt. 
For moderate driving strength the agreement is very good. }
\label{fig:b_x_correlation_ana_num}
\end{figure}
%
So, to calculate the two-time correlation function, one needs the 
steady state value of $ \rho_{ii}(\infty) $ and the time-dynamics 
of $ \rho^{jj}_{kk}(\tau) $ with the initial condition of 
$ \rho^{jj}_{kk}(0)=|\bra{k}\ket{j}|^2$.
To be explicit, for the biexciton-exciton correlation, we have
$ i=B,j=V,k=V,l=G$.
Therefore, the observable can be expressed as:
\begin{align}
g^{(2)}_{BVVG}(\tau) =& 
\lim\limits_{t \to \infty}
\frac{\ew{c^\dg_{BV}(t)c^\dg_{VG}(t+\tau)c^\ndg_{GV}(t+\tau)c^\ndg_{VB}(t)}}
{\ew{c^\dg_{BV}(t)c^\ndg_{VB}(t)}\ew{c^\dg_{VG}(t)c^\ndg_{GV}(t)} }
= 
\frac{\rho_{BB}(\infty)\rho^{VV}_{VV}(\tau)} 
{\rho_{BB}(\infty)\rho_{VV}(\infty) } 
= 
\frac{\rho^{VV}_{VV}(\tau)}{\rho_{VV}(\infty)}.
\end{align}
So, we just need the steady state values, and the dynamics for the 
$V-$exciton state with the initial conditions of $ \rho_{VV}=1$.
The exciton dynamics reads:
\begin{align}
\rho^{VV}_{VV}(\tau) 
=&
\frac{1}{2}
e^{-\Gamma\tau}
\left[
1
+
\frac{\Omega^2}{\Omega^2+\Gamma^2}
\left( 
\cosh(\Gamma \tau)
+
\frac{\Gamma^2}{\Omega^2}
\cos(\Omega \tau)
\right)
\right].
\end{align}
with the steady state 
$ \rho^{VV}_{VV}(\infty)=(\Omega/2)^2/(\Omega^2+\Gamma^2)$, approximately
$ 1/4 $ in the strong driving limit. 
The measured correlation, when the biexciton photon is detected first, 
reads:
\begin{align}
g^{(2)}_{BVVG}(\tau) 
=&
\frac{\rho^{VV}_{VV}(\tau)}{\rho_{VV}(\infty)}
=
2
e^{-\Gamma\tau}
\left(
1
+
\cosh(\Gamma \tau)
+
\alpha
[1+\cos(\Omega \tau)]
\right) \label{eq:g2_b_x},
\end{align}
with $\alpha=\Gamma^2/\Omega^2$.
In the long-time limit, the correlation always converges to $1$,
as it should be.
This can be seen, by taking into account, that 
$ \cosh(\Gamma\tau)=[\exp(\Gamma\tau)+\exp(-\Gamma\tau)]/2$.
Furthermore, the correlation starts always with
\begin{align}
g^{(2)}_{BVVG}(0) 
=&
4
\left(
1+
\alpha
\right), 
\end{align}
which reaches $4$ in the strong driving limit, i.e. $\alpha\rightarrow 0$.
The correlation cannot start with a smaller value.
For weaker driving, the initial correlation can be arbitrary 
high with $\alpha \gg 1 $.
The correlation function therefore always resides between 
$ 4 <g^{(2)}_{BVVG}(0)<\infty$.
This corresponds to the physical intuition, that the probability
to measure an exciton photon is dramatically higher if a biexciton
photon is detected.
The opposite is the case, if the exciton photon is measured first.
This is the next case, we are now considering.
If the exciton photon is detected first, then the correlation 
has the following index set $ i=V,j=G,k=B,l=V$ with the observable
of
\begin{align}
g^{(2)}_{VGGB}(\tau) =& 
\lim\limits_{t \to \infty}
\frac{\ew{c^\dg_{VG}(t)c^\dg_{BV}(t+\tau)c^\ndg_{VB}(t+\tau)c^\ndg_{GV}(t)}}
{\ew{c^\dg_{VG}(t)c^\ndg_{GV}(t)}\ew{c^\dg_{BV}(t)c^\ndg_{VB}(t)} }
=
\frac{\rho_{VV}(\infty)\rho^{GG}_{BB}(\tau)} 
{\rho_{BB}(\infty)\rho_{VV}(\infty) } 
=
\frac{\rho^{GG}_{BB}(\tau)}{\rho_{BB}(\infty)} .
\end{align}
For this sequence of detection events, the biexciton population 
dynamics needs to be calculated with following initial 
conditions $B(t)=0$, $\rho^{GG}_{BB}(0)=0$ and $D(t)=-1$.
The solution is
\begin{align}
\rho^{GG}_{BB}(t) 
=&
\frac{e^{-\Gamma t}\Omega^2}{2(\Omega^2+\Gamma^2)}
\left[
\sinh(\Gamma t)
-
\cos(\Omega t)
+
e^{-\Gamma t}
\right] 
=
\frac{e^{-\Gamma t}\Omega^2}{2(\Omega^2+\Gamma^2)}
\left[
\cosh(\Gamma t)
-
\cos(\Omega t)
\right] 
\end{align}
with the steady state 
$ \rho^{GG}_{BB}(\infty)=(\Omega/2)^2/(\Omega^2+\Gamma^2)$,
and the corresponding normalized correlation function reads:
\begin{align}
g^{(2)}_{VGGB}(\tau) 
=& 
2
e^{-\Gamma \tau}
\left[
\cosh(\Gamma \tau)
-
\cos(\Omega \tau)
\right] 
=
1
+
e^{-\Gamma \tau}
\left(
e^{-\Gamma \tau}
-
2
\cos(\Omega \tau)
\right). \label{eq:g2_x_b}
\end{align}
This correlation always starts with $ g^{(2)}_{VGGB}(0)=0  $,
i.e. it takes always a finite time to detect after an exciton 
a biexciton photon.
The long time limit is  $ g^{(2)}_{VGGB}(\infty)=1 $.
Furthermore, the maximum of the correlation can be inferred from
$\exp(-\Gamma\tau)\approx1$ and a driving amplitude chosen such that
$ \cos(\Omega \tau)=-1$, then $g^{(2)}_{GVVB}(\tau)<4 $.
So, the exciton-biexciton correlations oscillates between 
$ 0\le g^{(2)}_{GVVB}(\tau) <4$.
In Fig.~\ref{fig:b_x_correlation_ana_num}, the numerical and analytical
solutions of the photon correlation function between the detection of 
the biexciton and exciton and vice versa is plotted.
Note, $g^{(2)}_{GVVB}(\tau)=g^{(2)}_{BVVG}(-\tau)$. 
The numerical solution (black, dashed line) 
is obtained by evaluation of the full density 
matrix equation without any further approximation than necessary to
obtain the master equation in Eq.~\eqref{eq:full_meq}.
The analytical solutions of Eq.~\eqref{eq:g2_b_x} and~\eqref{eq:g2_x_b} 
(orange, solid line) agrees well in this driving limit: 
$ \Omega_L\ll \Delta $.
Numerical evaluations show that the approximations for the 
analytical solutions hold up to an acceptable mistake until
$\Omega_L \le \Delta/3$ for the exciton-biexciton direction 
$\tau<0$.
For $ \tau>0 $, the dynamics is for strong driving dominated by the  
dissipative dynamics, as the driving suppresses any oscillations with 
the factor $\Gamma^2/\Omega^2$.
Since the approximations hold perfectly in the weak driving limit,
the solution for $ \tau>0 $, namely for the biexciton-exciton 
measurement sequence, holds numerically for every driving, 
as long as the master equations stays valid.
The plot clearly shows, that a $ g^{(2)}(\tau)-$function of smaller than $ 1 $
always shows that an exciton photon is detected first, and in contrast
a $ g^{(2)}(\tau)-$value of larger than $ 4 $, refers to a biexciton-exciton
photon detection order.
The driving strength steers a Rabi oscillation signal between the exciton
and biexciton detection for $ \tau<1 $.
The stronger the driving, the faster the biexciton photon can be detected.
But there is no driving strength, where (theoretically) the $ g^{(2)}(\tau)-$
function does not initially start with value of zero. 
Experimentally, the time resolution may lead to finite values, but 
always to values smaller than $ 1 $, since the oscillations maximum 
amplitude is stronger damped than the oscillations minimum rises 
(biexciton decay versus exciton decay). 
%

\section{Dressed states} \label{sec:dressed_state_basis}
To enable a vanishing time-ordering, a spectral selection of 
the photons is necessary.
Until now, the calculations have not distinguished between an exciton
photon in the strong and weak driving limit.
Basically, three photons with slightly different frequencies 
can be emitted from the exciton, as well as from the 
biexciton state.
To unravel these three possibilities, it is necessary to 
change the basis of the theoretical description into 
the dressed state coordinates. 
In order to investigate the different spectral contributions and 
the dependence of the time-ordering on the driving strength, 
we express the dressed states in terms of the bare states and employ 
the solutions of the previous sections.
To address the dressed states individually, we find the Eigenstates
of the Hamiltonian, i.e. the Eigenstates of the coherent evolution only.
\begin{align}
H_R =& 
\Delta ( \ketbra{H}{H} + \ketbra{V}{V}  )
+ \Omega_L
( \ketbra{G}{H} + \ketbra{B}{H} + \text{h.a.}) 
\rightarrow 
\begin{pmatrix}
 0 & \Omega_L & 0 & 0 \\
\Omega_L & \Delta & \Omega_L & 0 \\
0 & \Omega_L & 0 & 0 \\
0 & 0 & 0 & \Delta 
\end{pmatrix}
, 
\end{align}
ordered in the basis 
$ \left\lbrace\ket{G},\ket{H},\ket{B},\ket{V}\right\rbrace $ 
and for the case of resonant two-photon driving of the biexciton and 
a polarization-selective driving of the horizontal exciton level,
without affecting the vertical-polarized transitions.
The diagonalization results in four Eigenvalues
\begin{align}
e_0 &=0, \quad
e_1=\Delta, \quad
e_3= 
\frac{1}{2} 
\left(
\Delta + \sqrt{8\Omega^2 + \Delta^2}
\right), \quad
e_4 = 
\frac{1}{2} 
\left(
\Delta - \sqrt{8\Omega^2 + \Delta^2}
\right),
\end{align}
with the corresponding Eigenvectors:
\begin{align}
\ket{+} 
=& \frac{\Omega}{\sqrt{2\Omega^2+e^2_3}}
\left( \ket{G} + \frac{e_3}{\Omega} \ket{H} + \ket{B} \right)
,\quad \ket{V} = \ket{V} \\
\ket{-} 
=& \frac{\Omega}{\sqrt{2\Omega^2+e_4^2}}
\left( \ket{G} + \frac{e_4}{\Omega} \ket{H} + \ket{B} \right)
,\quad \ket{0} = \frac{1}{\sqrt{2}} \left( \ket{B}-\ket{G} \right) . 
\end{align}
Orthogonality and normalization can be proven via helpful 
relation between the eigenvalues, cf.~\ref{app:symplectic_factors}.
From the Eigenvalues it can be already seen, that the biexciton,
the $H-$exciton and the ground state form a superposition due to 
the coherent driving.
The external laser field creates coherences in between these states,
which leads to three frequency-differentiated biexciton-exciton 
transitions $ \sig{+}{V}, \sig{-}{V}, \sig{0}{V} $ 
and also three exciton-ground state transitions
$\sig{V}{+}, \sig{V}{-}, \sig{V}{0} $.
If the driving is weak, the frequencies do not appear as separate
peaks for they lie all within the radiative linewidth $ \Gamma$.
But in the strong driving limit, Mollow physics appear and become 
spectrally resolved in the spectrum.
There is a particularly interesting feature that becomes visible in the
strong driving limit.
In contrast to the full exciton-biexciton correlation, where at least
in one delay direction, a $ g^{(2)}-$value smaller than $1$ always 
occurs, it is possible to create a signal that never exhibits a
$ g^{(2)}-$value smaller than one. 
This is rendered possible via a decay in form of a superposition and is 
a specific property of the strong driving limit, where the coherences 
are strongly enhanced.
Those superpositions do not differentiate between exciton and 
biexciton photons anymore.
In this case, the time-ordering in the cascade has been lifted.
And it is not possible to judge via the $ g^{(2)}-$signal whether a 
biexciton or exciton photon has been measured first.
In the following, we calculate analytically the signal of the 
correlation functions, which exhibits in the strong driving limit 
such a vanishing time-reordering.
The observable reads:
\begin{align}
g^{(2)}_{+VV+}(\tau) =& 
\lim\limits_{t \to \infty}
\frac{\ew{c^\dg_{+V}(t)c^\dg_{V+}(t+\tau)c^\ndg_{+V}(t+\tau)c^\ndg_{V+}(t)}}
{\ew{c^\dg_{+V}(t)c^\ndg_{V+}(t)}\ew{c^\dg_{V+}(t)c^\ndg_{+V}(t)} }.
\end{align}
The system undergoes the cascade either from 
$ \ket{+}\rightarrow\ket{V}\rightarrow\ket{+} $.
This observable will be calculated in the next section, and it will be 
shown that the $0-$state observable exhibits the same dynamics with
\begin{align}
g^{(2)}_{0VV0}(\tau) =& 
\lim\limits_{t \to \infty}
\frac{\ew{c^\dg_{0V}(t)c^\dg_{V0}(t+\tau)c^\ndg_{0V}(t+\tau)c^\ndg_{V0}(t)}}
{\ew{c^\dg_{0V}(t)c^\ndg_{V0}(t)}\ew{c^\dg_{V0}(t)c^\ndg_{0V}(t)} }
=g^{(2)}_{+VV+}(\tau).
\end{align}
For the experimental signal, in Sec.~\ref{sec:experiment},
another correlation function is of importance, i.e. a cross-correlation
between the dressed states:
\begin{align}
g^{(2)}_{+VV0}(\tau) =& 
\lim\limits_{t \to \infty}
\frac{\ew{c^\dg_{+V}(t)c^\dg_{V0}(t+\tau)c^\ndg_{0V}(t+\tau)c^\ndg_{V+}(t)}}
{\ew{c^\dg_{+V}(t)c^\ndg_{V+}(t)}\ew{c^\dg_{V0}(t)c^\ndg_{0V}(t)} }.
\end{align}
Due to intrinsic symmetric reasons in detail explained 
in~\ref{app:dressed_correlation}, this cross-correlation 
stays the same whether first  the $ 0-$state photon and 
the $ +-$state photon is detected or the other way round, 
expressed in a formula: 
$g^{(2)}_{+VV0}(\tau)=g^{(2)}_{0VV+}(\tau)$.
The $-$state photons are not included in the experimental
signal as those photons are too close to the frequency of the 
laser photons and cannot easily be distinguished from elastic
scattering events.
%

\section{Photon-Photon correlation of selected dressed states}
\label{sec:dressed_state_g2}
\begin{figure}[t!]
\centering
\includegraphics[width=16cm]{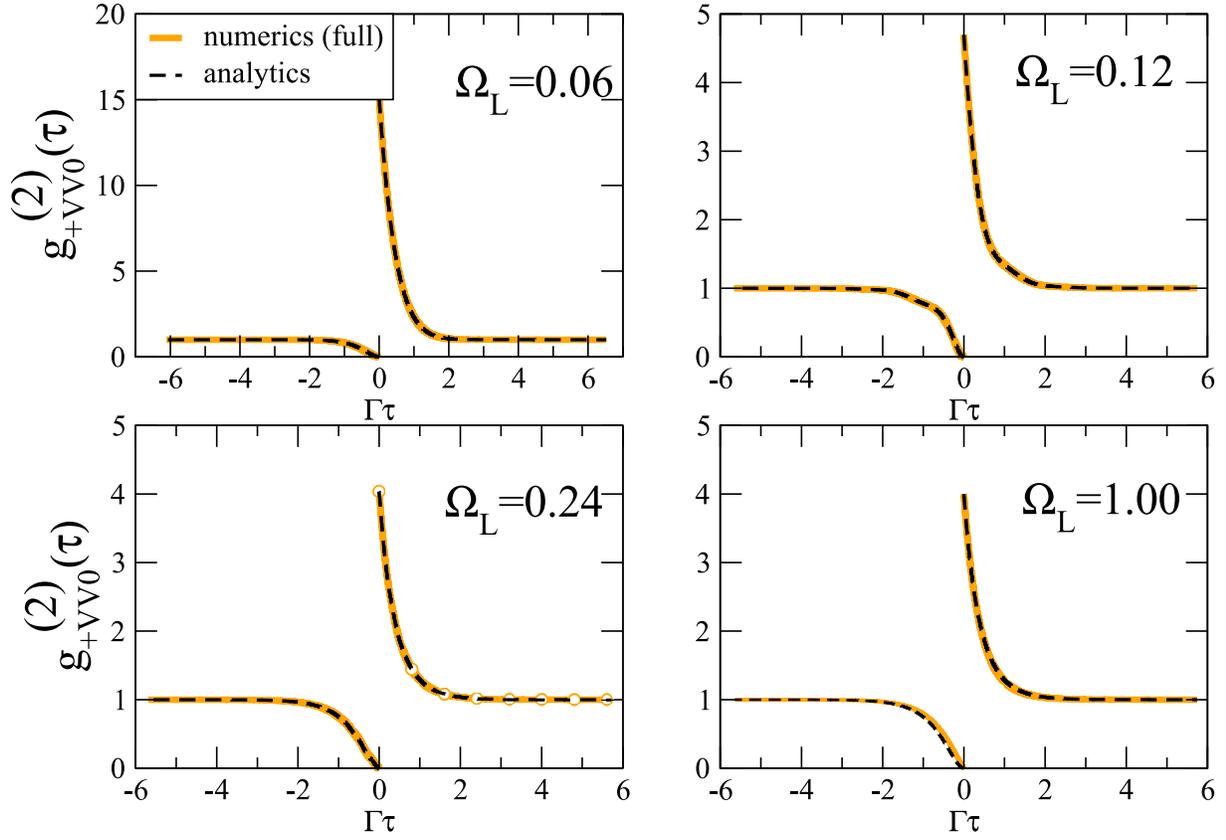}
\caption{Photon-photon correlations between $ +$ and 
$0$-state photons for different driving strengths 
$ \Omega_L[\text{ps}^{-1}]$ with 
$ \Gamma=0.002\text{ps}^{-1}$ and $ \Delta=3.0\text{ps}^{-1}$. The 
numerical solution is derived from the complete master equation. The 
analytical solutions are given in the text in the adiabatic limt. 
For moderate driving strength the agreement is very good. }
\label{fig:+_0_correlation_ana_num}
\end{figure}

%
The detection sequence is again two-fold, either the $ \ket{+} $
or the $ \ket{V} $ photon is detected first.
The detection is polarized selective, and so the calculation is
simplified due to a vanishing $ \ket{H} $ contribution.
Using the same method as before, we need the correlation
with following flip operators for $ \tau>0$: 
\begin{align}\notag
g^{(2)}_{+VV+}(\tau) =& 
\lim\limits_{t \to \infty}
\frac{
\ew{\sig{+}{V}(t)\sig{V}{+}(t+\tau)\sig{+}{V}(t+\tau)\sig{V}{+}(t)} 
}{\ew{\sig{+}{+}(t)}\ew{\sig{V}{V}(t)}}
\end{align}
Therefore, the observable can be expressed as:
\begin{align}
g^{(2)}_{+VV+}(\tau) =& 
\frac{\rho_{++}(\infty)\rho^{VV}_{VV}(\tau)} 
{\rho_{++}(\infty)\rho_{VV}(\infty) } 
=
2
e^{-\Gamma\tau}
\left(
1
+
\cosh(\Gamma \tau)
+
\alpha
[1+\cos(\Omega \tau)]
\right) . \label{eq:g+v}
\end{align}
This part of the correlation has not change in comparison to 
the previous calculation.
This is due to the fact, that the conditional probablity of 
the $ V-$exciton state is measured, and this state does not 
change with the driving strength, for it is decoupled from the 
laser.
However, for $ \tau<0 $, the correlation function reads:
\begin{align}
g^{(2)}_{V++V}(\tau) =& 
\frac{\rho_{VV}(\infty)\rho^{++}_{++}(\tau)} 
{\rho_{VV}(\infty)\rho_{++}(\infty) } 
= 
\frac{\rho^{++}_{++}(\tau)}{\rho_{++}(\infty)}.
\end{align}
We need to express the dressed state basis in terms of the 
bare state to use the aforementioned solution.
Under the condition, that the horizontal-polarized photons 
are not detected, the corresponding dressed state reads:
\begin{align}
\ket{+} 
=& \frac{1}{\sqrt{2}}
\left( \ket{G} + \ket{B} \right).
\end{align}
Now, the density matrix dynamics needs to be expressed in 
terms of this superposition:
\begin{align}   \notag
g^{(2)}_{V++V}(\tau) =& 
\text{Tr}
\left[ 
\ketbra{+}{+}
\frac{\sig{+}{+}
(\tau)}{\rho_{++}(\infty)}
\right]
=
\text{Tr}
\left[ 
\left(
\ketbra{B}{B}
+
\ketbra{B}{G}
+
\ketbra{G}{B}
+
\ketbra{G}{G}
\right)
\frac{\sig{+}{+}
(\tau)}{2\rho_{++}(\infty)}
\right]
.
\end{align}
%
\begin{figure}[t!]
\centering
\includegraphics[width=16cm]{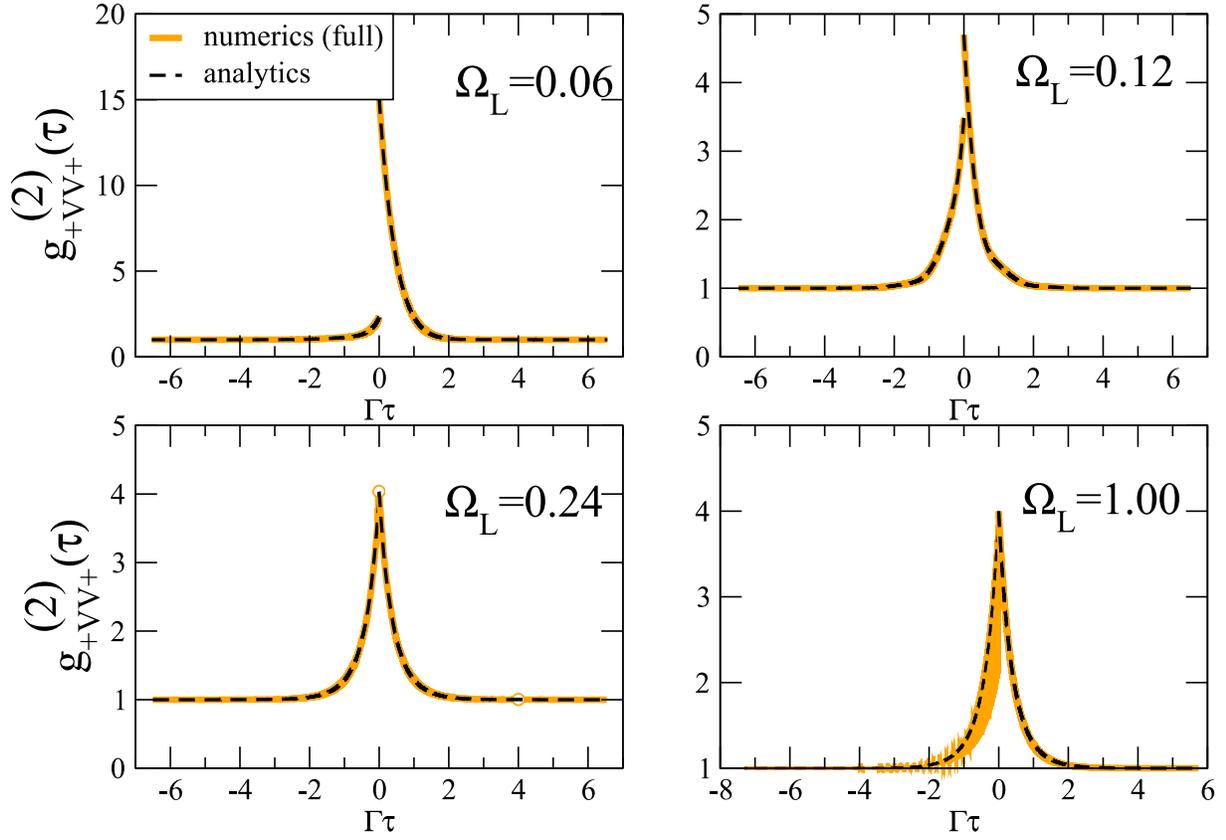}
\caption{Photon-photon correlations between $+$ biexciton and 
$+$ exciton photon for different driving strengths 
$ \Omega_L[\text{ps}^{-1}]$ with 
$ \Gamma=0.002\text{ps}^{-1}$ and $ \Delta=3.0\text{ps}^{-1}$. The 
numerical solution is derived from the complete master equation. The 
analytical solutions are given in the text in the adiabatic limt. 
For moderate driving strength the agreement is very good. }
\label{fig:+_+_correlation_ana_num}
\end{figure}
%
Due to the superposition, the calculation can be more tedious
than before, if done separately.
However, it is more feasible to calculate the dynamics 
directly via a superpostion of the initial conditions. 
After measuring a $V-$photon, the system is in the 
following mixture of states:
$ \rho^{++}(0)
=
\left(
\ketbra{B}{B}
+
\ketbra{B}{G}
+
\ketbra{G}{B}
+
\ketbra{G}{G}
\right)/2$.
Therefore, the initial conditions read:
\begin{align}
D^{++}_0 &:=\rho^{++}_{BB}-\rho^{++}_{GG}=0 \\
B^{++}_0 &:=\rho^{++}_{BG}-\rho^{++}_{GG}=0 \\
\Sigma^{++}_0&:=\rho^{++}_{BB}+\rho^{++}_{VV}+\rho^{++}_{HH}+\rho^{++}_{GG}=2.
\end{align}
The correlation itself reads with its corresponding initial condition:
\begin{align}
\ew{\sig{+}{+}(\tau)}|_{\rho^{++}(0)} 
=&
\frac{1}{4}
\left(
\rho^{++}_{BB}(\tau)
+
\rho^{++}_{GG}(\tau) 
+
2
\text{Re}
\left[
\rho^{++}_{GB}(\tau) 
\right]
\right).
\end{align}
Given all the initial conditions, the correlation can easily 
be computed, cf.~\ref{app:dressed_correlation},
and reads:
\begin{align}
g^{(2)}_{V++V}(\tau) =&
1
+
2\frac{1+\alpha}{1+2\alpha}
\left[
e^{-2\Gamma\tau}
\left(
1-
\frac{1}{2}
\frac{1}{1+\alpha}
\right)
+
e^{-\Gamma\tau}
\left(
1-
\frac{\alpha\cos(\Omega\tau)}{1+\alpha}
\right)
\right] \label{eq:gv+},
\end{align}
with $ \alpha :=\Gamma^2/\Omega^2 $.
This formula constitutes the main result of this paper.
In Fig.~\ref{fig:+_+_correlation_ana_num}, 
the full numerical solution without adiabatic approximation 
(solid, orange line) is compared with the analytical
formula (dashed, black line) given in Eq.~\eqref{eq:g+v} and~\eqref{eq:gv+}.
The results agree and confirm the analytical calculation up to 
driving strength $ \Omega_L<\Delta/3 $.
For very strong driving, fast oscillations appear below the 
adiabatic curve, see Fig.~\ref{fig:+_+_correlation_ana_num}
(lower right panel).
In this limit, the fast oscillations between the exciton states
lead even to values of below $ 1 $.
However, the initial value of the correlations functions is always 
larger than $ 1 $, as the analytical formula shows:
\begin{align}
g^{(2)}_{V++V}(0)
&=
4\frac{1+\alpha}{1+2\alpha}.
\end{align}
So, for large driving compared to the decay constant,
i.e. $ \alpha \rightarrow 0$, the initial value
approaches $g^{(2)}_{V++V}(0)\rightarrow4$, showing exactly the 
time-reordering reversal.
The detection schemes are not distinguishable anymore,
since $g^{(2)}_{V++V}(0) \approx g^{(2)}_{+VV+}(0)$ is valid.
This is the specific result of such a two-photon driving 
and frequency-polarization filtered detection setup.
In the low excitation regime $ \alpha \gg 1 $ , it can be seen 
that the minimum of the correlation function is $ 2 $.
Investigating the total maximum of the correlation by assuming 
$ \cos(\Omega\tau)=-1$, the correlation in the adiabatic regime 
stays with the interval $ 2< g^{(2)}_{V++V}(0)<4$.
All these properties can be seen in the 
Fig.~\ref{fig:+_+_correlation_ana_num} for the different driving 
strength.
The cross-correlation between the dressed states read:
\begin{align}\notag
g^{(2)}_{+VV0}(\tau) =& 
\lim\limits_{t \to \infty}
\frac{
\ew{\sig{+}{V}(t)\sig{V}{0}(t+\tau)\sig{0}{V}(t+\tau)\sig{V}{+}(t)} 
}{\ew{\sig{+}{+}(t)}\ew{\sig{V}{V}(t)}}
=
\frac{\rho_{++}(\infty)\rho^{VV}_{VV}(\tau)} 
{\rho_{++}(\infty)\rho_{VV}(\infty) } 
=
\frac{\rho^{VV}_{VV}(\tau)} 
{\rho_{VV}(\infty) }, 
\end{align}
which we have already calculated. 
However, the inverted correlation cannot be expressed with the 
derived results from before, see for 
details~\ref{app:dressed_correlation}:
\begin{align}
g^{(2)}_{V0+V}(\tau) =& 
\lim\limits_{t \to \infty}
\frac{
\ew{\sig{V}{0}(t)\sig{+}{V}(t+\tau)\sig{V}{+}(t+\tau)\sig{0}{V}(t)} 
}{\ew{\sig{+}{+}(t)}\ew{\sig{V}{V}(t)}}
=
\frac{\rho^{00}_{++}(\tau)} 
{\rho_{++}(\infty) }.
\end{align}
The solution differs just in one sign and reads:
\begin{align}
g^{(2)}_{V+0V}(\tau) =&
1
+
2\frac{1+\alpha}{1+2\alpha}
\left[
e^{-2\Gamma\tau}
\left(
1-
\frac{1}{2}
\frac{1}{1+\alpha}
\right)
-
e^{-\Gamma\tau}
\left(
1+
\frac{\alpha\cos(\Omega\tau)}{1+\alpha}
\right)
\right]  \label{eq:gv+0v} \\
=&
1
+
\frac{1}{1+2\alpha}
\left[
e^{-2\Gamma\tau}
\left(1+2\alpha\right)
-
2
e^{-\Gamma\tau}
\left(
1+\alpha+\alpha\cos(\Omega\tau)
\right)
\right] .
\end{align}
In contrast to the other dressed correlation in this 
section, this one starts always with $ 0 $ for $ \tau<0 $, cf. 
Fig.~\ref{fig:+_0_correlation_ana_num}. 
%
%
The antibunching feature is independent of the driving 
strength.
The bunching on the biexciton-exciton side $ \tau>0 $ is
however decreased for stronger driving and the numerical
and analytical solutions agree well.
With the symmetries at hand, we can now turn to the experimental
data for different intensities and compare the theoretical 
findings with the experiments.
%

\section{Comparison with the experiment}
\label{sec:experiment}
\begin{figure}[t!]
\centering
\includegraphics[width=16cm]{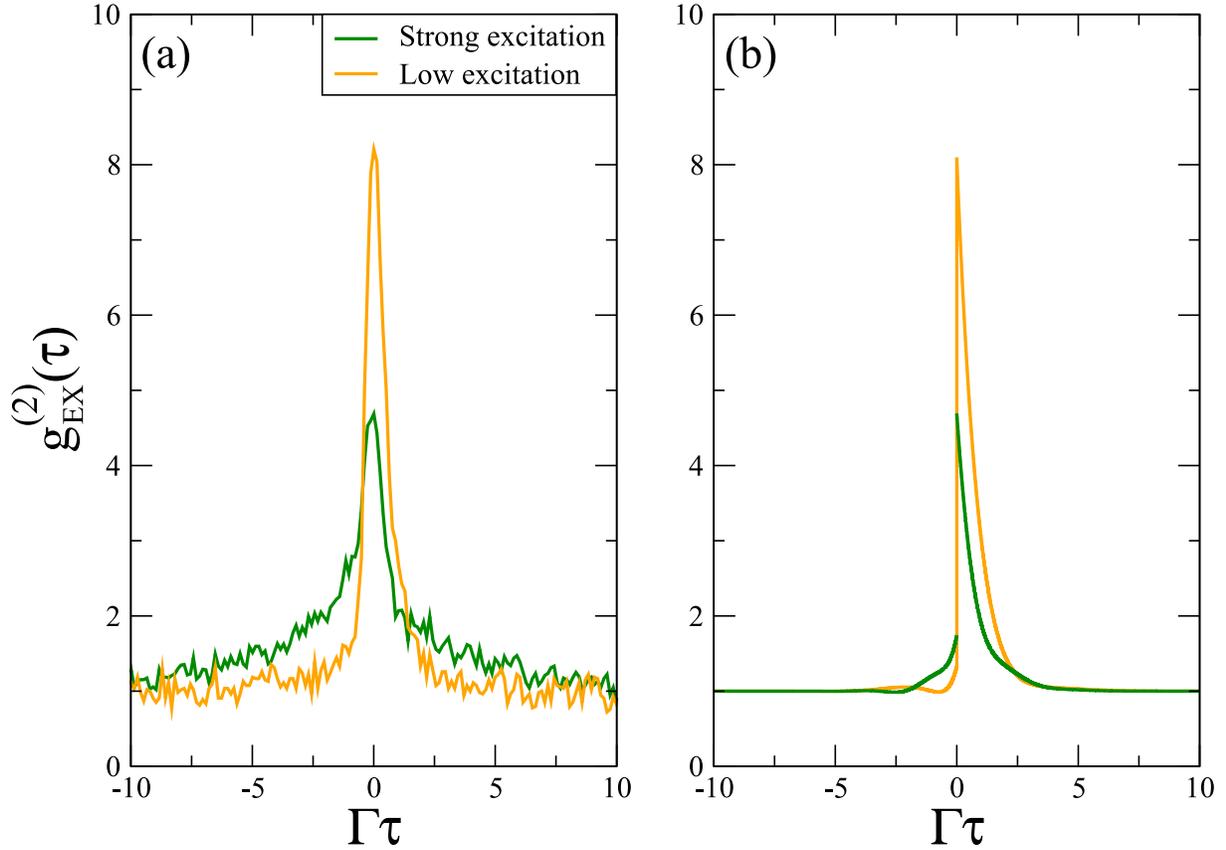}
\caption{Comparison between experiment (a) and theory (b) for two low
$(30\mu\text{W})$ and strong excitation $(100\mu\text{W})$ power
and the exciton-biexciton photon-photon correlations without 
$ \ket{-}$ state contributions. The qualitative agreement is 
very good. Both plots show the symmetrization of detection events 
for stronger excitation (green) in comparison to lower excitation
(orange). The experimental data is shown with a substracted 
background, in particular for the lower excitation.}
\label{fig:exp_theo_comp}
\end{figure}

In the experiment, the biexciton-exciton correlation is taken 
only partially.
The contribution from the $ \ket{-}$-state are difficult to 
distinguish from the laser photons and are therefore excluded
from the detection setup.
So, the biexciton-exciton correlation reads in the experiment:
\begin{align}
g^{(2)}_\text{EX}(\tau) 
=&
\frac{1}{4}
\left(
g^{(2)}_\text{+VV+}(\tau)
+
g^{(2)}_\text{0VV0}(\tau)
+
g^{(2)}_\text{+VV0}(\tau)
+
g^{(2)}_\text{0VV+}(\tau)
\right),
\end{align}
which are all possible detection events for this chosen frequency-window.
We take advantage of our definition: 
$ g^{(2)}_\text{ijkl}(\tau)=\rho^{jj}_{kk}(\tau)/\rho^{jj}_{kk}(\infty)$,
i.e. the correlation describes the conditional probability to measure 
the photon from state $ \ket{k} $, after a photon is detected, which 
collapses the state of the system to the state $\ket{j}$.
Using this definition, we can write the biexciton-exciton detection
sequence as:
\begin{align}
g^{(2)}_\text{EX}(\tau) 
=&
\frac{\rho^{VV}_{VV}(\tau)}{\rho^{VV}_{VV}(\infty)}
=
3
+
e^{-2\Gamma\tau}
+
2
\alpha
e^{-\Gamma\tau}
[1+\cos(\Omega \tau)]
.
\end{align}
However, the exciton-biexciton direction is more complicated.
All possible detection events read:
\begin{align}
g^{(2)}_\text{EX}(\tau) 
=&
\frac{1}{4}
\left(
g^{(2)}_\text{V++V}(\tau)
+
g^{(2)}_\text{V00V}(\tau)
+
g^{(2)}_\text{V+0V}(\tau)
+
g^{(2)}_\text{V0+V}(\tau)
\right),
\end{align}
but we know, out of symmetry reasons that 
$ \rho^{++}_{++}(\tau)=  \rho^{00}_{00}(\tau)$ and 
$ \rho^{++}_{00}(\tau)=  \rho^{00}_{++}(\tau)$.
Therefore, we can write:
\begin{align}
g^{(2)}_\text{EX}(\tau) 
=&
\frac{1}{2}
\left(
\frac{\rho^{++}_{++}(\tau)}{\rho^{++}_{++}(\infty)}
+
\frac{\rho^{++}_{00}(\tau)}{\rho^{++}_{00}(\infty)}
\right)
=1+e^{-2\Gamma\tau}
-\frac{2\alpha\cos(\Omega\tau)}{1+2\alpha}e^{-\Gamma\tau},
\end{align}
so for $ \tau\rightarrow0 $, the conditional probability
reads
\begin{align}
g^{(2)}_\text{EX}(0) 
=&
1+1
-\frac{2\alpha}{1+2\alpha}
=
2-\frac{2\alpha+1-1}{1+2\alpha}
=
1+\frac{1}{1+2\alpha} >1.
\end{align}
Both detection sequences show bunching. 
In Fig.~\ref{fig:exp_theo_comp}(a), the bunching effect is shown 
in the experimental data for the strong (green) and weak driving 
regime (orange).
For stronger excitation the detection around $ \tau=0 $ symmetrizes,
which signifies a superposition in detection events rendering the 
detection order partially indistinguishable.
However, in contrast to the case of $ g^{(2)}_\text{+VV+}(\tau)$ 
correlation function, the peak around $ \tau=0 $ cannot be 
completely symmetrical due to antibunching contribution from the 
cross-correlation, e.g. $ g^{(2)}_\text{+VV0}(\tau)$.
The analytical theory reproduces the effect well, cf. 
Fig.~\ref{fig:exp_theo_comp}(b).
Note, the theory is not convoluted with the detector-response
function, as the focus of this work is to provide general formulas
for the correlation functions and not to discuss detailed experimental
data.
In consequence, the comparison to the experiment is only 
qualitatively without fitting the decay and excitation 
constant properly.
Nevertheless, the theory reproduces the important fact, that
antibunching is highly unlikely in this dectection setup.
For antibunching is created in a superposition state of all
occuring dressed state emission events, locked, phase-matched
to the exciton-emisson. 
The signal is not just a sum of the correlation functions but 
a constructive and destructive interference between photon detections.
Leading to the conclusion, that it is much harder in this setup
to render the antibunching visible out of the aforementioned reasons.
%

\section{Conclusion}\label{sec:conclusion}
We have calculated analytically the two-time correlations of 
the exciton-biexciton cascade in the adiabatic limit, where
the detuning between the laser and the excitonic transition
is much larger than the driving amplitude.
In this limit, we discussed the correlation of the biexciton-
exciton photons, which showed clearly a time-reordering.
It is always possible to distinguish the detection sequence
by their $ g^{(2)}_{BVVG}(\tau)$ values around zero 
$ -1/\Gamma\ll\tau\ll1/\Gamma $, i.e. $g^{(2)}(\tau)_{BVVG}>1$ 
the biexciton has been detected, then the exciton, and 
$g^{(2)}_{XB}(\tau)<1$ vice versa.
If the dressed states can spectrally be selected, it is 
possible to erase this time-ordering.
We showed this analytically by calculating the correlation
function in the dressed state basis $ g^{(2)}_{+VV+}(\tau)$.
Furthermore, we showed with our calculations the limits 
of the correlation functions in the idealized situation of 
an isolated four level system, and why it is difficult to
observe in such a system the antibunching effect.

\ \newline
The research leading to these results has received funding from
the European Research Council (ERC) under the European Union's Seventh 
Framework ERC Grant Agreement No. 615613 and from the German Research
Foundation (DFG) through SFB 787 via Projects No. RE2974/4-1 and 
No. RE2974/12-1.
A.C. gratefully acknowledges support from the
SFB 910: ``Control of self-organizing nonlinear systems.``
%
\begin{appendix}

\section{Comparison to the two-level case}\label{app:two_level_compare}
In this section, we show that the resonant two-photon driven
four-level system obeys the well-known two-level physics.
The resonant Mollow problem is cast into the following master equation:
\begin{align}
\dot \rho =& -i \left[\Omega_2(\ketbra{G}{B}+\ketbra{B}{G}),\rho\right] 
+ \Gamma_2
\mathcal{D}[\sig{G}{B}]
\rho, \label{eq:two_level_meq}
\end{align}
using the standard Lindblad form 
$\mathcal{D}[J]\rho=2 J\rho J^\dg-\lbrace J^\dg J,\rho\rbrace$
and denoting the parameter with $_2$ to distinguish them 
from the four-level case.
Using the master equation, the follow set of differential equations of
motion is yield with $ \rho_{ij}=\bra{i}\rho\ket{j}$:
\begin{align}
\dot \rho_{GG} 
=&
2 \Gamma_2 \rho_{BB}
+i\Omega_2
\left( 
\rho_{BG}
-
\rho_{GB}
\right)
=
2 \Gamma_2 (\Sigma_{2,0}-\rho_{GG})
+i\Omega_2
\left( 
\rho_{BG}
-
\rho_{GB}
\right), \\
\dot \rho_{BG} 
=&
-
\Gamma_2 
\rho_{BG}
-i
\Omega_2
\left( 
\rho_{BB}
-
\rho_{GG}
\right), \\
\dot \rho_{BB} 
=&
-2 \Gamma_2 \rho_{BB}
-i\Omega_2
\left( 
\rho_{BG}
-
\rho_{GB}
\right),  
\end{align}
where we used again $\rho_{BB}+\rho_{GG}=\Sigma_{2,0}$
with typically $\Sigma_{2}(t=0)=\Sigma_{2,0}=1$.
We substract the population dynamics from each other 
and also the polarization dynamics:
$ D(t) := \rho_{BB}(t) - \rho_{GG}(t)$
and the linear independent imaginary part of the polarization
between both states
$B(t) := \rho_{BG}(t) - \rho_{GB}(t) $:
\begin{align}
\dot B(t)
=&
-
\Gamma_2 
\rho_{BG}
+
\Gamma_2 
\rho_{GB}
-i2
\Omega_2
\left( 
\rho_{BB}
-
\rho_{GG}
\right)
=
-\Gamma_2 B(t)
-i2\Omega_2 D(t)
\notag , \\
\dot D(t)=&
-2 \Gamma_2 \rho_{BB}
-2 \Gamma_2 (\Sigma_{2,0}-\rho_{GG})
-i2\Omega_2
\left( 
\rho_{BG}
-
\rho_{GB}
\right)
=
-2\Gamma_2\left( D(t)+\Sigma_{2,0}\right)
-i2\Omega_2 B(t)
\notag.  
\end{align}
This is exactly the same set of equation of motions as
the adabiatic version of the two-photon biexciton driving
case. 
The difference is hidden in $ \Sigma_{2,0} \neq \Sigma_0 $ 
in the main text. 
The dynamics of the exciton level play a role and allow
different values.
Furthermore, the transition operator $ \sig{G}{B} $ 
does not correspond to a single-photon emission but in this 
case to a two-photon emission process. 
However, the well-known solution of the Mollow problem 
apply to the four-level case in the adiabatic limit, also,
but it needs to be unraveled in terms of single-photon 
detection events.
\section{Laplace solution}\label{app:laplace_solution}
In this section, the analytical solution of the inversion 
and transition amplitude dynamics are derived.
The problem is identical to the dynamics of a two-level system
but with a different dissipative dynamics underlying.
For completeness, we derive the solution via Lapace transformation.
In text books, mainly a diagonalization technique is applied \cite{scully}.
We transform Eq.~\eqref{eq:dt} and ~\eqref{eq:bt} 
the into the Laplace domain with 
$ \Sigma_0=\rho_{GG}(0)+\rho_{BB}(0)+\rho_{VV}(0)+\rho_{HH}(0) $:
\begin{align}
s\bar D
-
D(0)
=& 
-\Gamma \bar D - \Gamma \frac{\Sigma_0}{s}  - i\Omega \bar B \\
s\bar B
- B(0)
=& 
-\Gamma \bar B - i\Omega \bar D.
\end{align}
Let's solve this, and abbreviate $ B(0)=B_0 $ and $ D(0)=D_0 $.
We see that:
\begin{align}
\bar B =& 
\frac{B_0}{s+\Gamma}
-
i\Omega
\frac{\bar D}{s+\Gamma}
\end{align}
We insert this solution into the equation for the inversion:
\begin{align}
s\bar D
-
D_0
=& 
-\Gamma \bar D - \Sigma_0\frac{\Gamma}{s}  
- i\Omega 
\left(
\frac{B_0}{s+\Gamma}
-
i\Omega
\frac{\bar D}{s+\Gamma}
\right) \\
s\bar D
=& 
D_0
-\Gamma \bar D - \Sigma_0\frac{\Gamma }{s}  
- i\Omega 
\frac{B_0}{s+\Gamma}
-
\Omega^2
\frac{\bar D}{s+\Gamma} \\
\bar D
=& 
D_0
\frac{(s+\Gamma)}{(s+\Gamma)^2+\Omega^2}
-i\Omega
B_0
\frac{1}{(s+\Gamma)^2+\Omega^2}
-
\Sigma_0
\frac{\Gamma}{s}  
\frac{s+\Gamma}{(s+\Gamma)^2+\Omega^2}
\end{align}
and for the polarisation:
\begin{align} 
\bar B =& 
\frac{B_0}{s+\Gamma}
-
\frac{i\Omega\bar D}{s+\Gamma} 
=
\frac{B_0(s+\Gamma)}{(s+\Gamma)^2+\Omega^2}
-
i\Omega D_0
\frac{i\Omega D_0}{(s+\Gamma)^2+\Omega^2}
+
\frac{\Gamma}{s}  
\frac{\Sigma_0 i\Omega}{(s+\Gamma)^2+\Omega^2}.
\end{align}
We use following Laplace transform identities:
\begin{align}
\bar f(s)=
\frac{1}{(s+\Gamma)^2+\Omega^2}
\longrightarrow &
f(t)=
\frac{1}{\Omega}
e^{-\Gamma t} \sin(\Omega t) \\
\bar f(s)=
\frac{s+\Gamma}{(s+\Gamma)^2+\Omega^2}
\longrightarrow &
f(t)=
e^{-\Gamma t} \cos(\Omega t)
\\
\bar f(s)=
\frac{1}{s}
\frac{1}{(s+\Gamma)^2+\Omega^2}
\longrightarrow &
f(t)=
\frac{1}{\Omega^2+\Gamma^2}
-
\frac{e^{-\Gamma t}}{\Omega}
\frac{\Omega\cos(\Omega t)+\Gamma\sin(\Omega t)}{\Omega^2+\Gamma^2}
 \notag
\\
\bar f(s)=
\frac{1}{s}
\frac{s+\Gamma}{(s+\Gamma)^2+\Omega^2}
\longrightarrow &
f(t)=
\frac{\Gamma}{\Omega^2+\Gamma^2}
+
e^{-\Gamma t}
\
\frac{\Omega\sin(\Omega t)-\Gamma\cos(\Omega t)}{\Omega^2+\Gamma^2}
\end{align}
In the time domain, the inversion dynamics read
\begin{align}
D(t)
=& 
D_0
e^{-\Gamma t} \cos(\Omega t)
-iB_0
e^{-\Gamma t} \sin(\Omega t)
\\ &
-\Sigma_0\Gamma 
\left[
\Gamma_n
+
e^{-\Gamma t}
\
\left(
\Omega_n\sin(\Omega t)-\Gamma_n\cos(\Omega t)
\right)
\right]
\end{align}
which fulfills: $ D(t=0)=D_0 $
with abbreviated normalized constants 
\begin{align}
\Omega_n =& \frac{\Omega}{\Omega^2+\Gamma^2} 
\qquad
\Gamma_n = \frac{\Gamma}{\Omega^2+\Gamma^2} .
\end{align}
The general equation of polarisation reads:
\begin{align}
B(t) 
=& 
B_0 \
e^{-\Gamma t} \cos(\Omega t)
-
iD_0 \
e^{-\Gamma t} \sin(\Omega t)
\\
&
+
\Sigma_0
i\Gamma
\left[
\Omega_n
-
e^{-\Gamma t}
\left(
\Omega_n\cos(\Omega t)+\Gamma_n\sin(\Omega t)
\right)
\right].
\end{align}
Given the initial conditions and the probability conservation,
the dynamics of all other quantities can be calculated from these 
two solutions.
\section{Explicit solution of the biexciton dynamics}
\label{app:explicit_general_solutions}
To derive the solution for the biexciton dynamics, we need 
to use the transition dynamics $ B(t) $
and integrate its equation of motion.
As we know the equation of motion of the biexciton density, 
we yield:
\begin{align}
\dot \rho_{BB} 
=&
-2\Gamma
\rho_{BB}
+
\Omega
\text{Im}
\left[ 
\rho_{BG}
\right]
=
-2\Gamma
\rho_{BB}
+
\frac{\Omega}{2i}
\left[ 
\rho_{BG}
-
\rho_{GB}
\right]
=
-2\Gamma
\rho_{BB}
+
\frac{\Omega}{2i}
B(t)
.
\end{align}
So we can write
\begin{align}
\rho_{BB}(t) 
=&
\rho_{BB}(0) 
e^{-2\Gamma t}
+
\frac{\Omega}{2}
\rho^I_{BB}(t) .
\end{align}
with the inhomogeneous solution,
that we need to calculate
\begin{align}
\rho^I_{BB} 
=&
e^{-2\Gamma t}
\int_0^t
dt_1
e^{2\Gamma t_1}
2\text{Im}\left[\rho_{BG}\right]
=
-i
e^{-2\Gamma t}
\int_0^t
dt_1
e^{2\Gamma t_1}
\
B(t_1) 
\\
=&
-i
\left(
B_0 
-\Sigma_0i\Gamma\Omega_n
\right)
e^{-2\Gamma t}
\text{Re}
\left[
\int_0^t
dt_1
e^{(\Gamma+i\Omega) t_1} \
\right]\\
& -
\left(
\Sigma_0
\Gamma\Gamma_n
+
D_0 
\right)
e^{-2\Gamma t}
\text{Im}
\left[
\int_0^t
dt_1
e^{(\Gamma+i\Omega) t_1} \
\right]
+
\Sigma_0
\Gamma
\Omega_n
\frac{1-e^{-2\Gamma t}}{2\Gamma}
. 
\end{align}
The integrals are evaluated:
\begin{align}
\rho^I_{BB}(t)=& \notag
-i
\left(
B_0 
-i\Sigma_0\Gamma\Omega_n
\right)
e^{-2\Gamma t}
\text{Re}
\left[
\frac{\Gamma-i\Omega}{\Omega^2+\Gamma^2}
\left(
e^{(\Gamma+i\Omega) t}
-1
\right)
\right] 
\\
&
-
\left(
\Sigma_0
\Gamma\Gamma_n
+
D_0 
\right)
e^{-2\Gamma t}
\text{Im}
\left[
\frac{\Gamma-i\Omega}{\Omega^2+\Gamma^2}
\left(
e^{(\Gamma+i\Omega) t}
-1
\right)
\right]
+
\Sigma_0
\Gamma
\Omega_n
\frac{1-e^{-2\Gamma t}}{2\Gamma}
.
\end{align}
Using
\begin{align}
\text{Re}
\left[
(\Gamma-i\Omega)
(e^{i\Omega t}-e^{-\Gamma t})
\right]
=&
\Gamma\cos(\Omega t)+\Omega\sin(\Omega t)
-\Gamma
e^{-\Gamma t}
\\
\text{Im}
\left[
(\Gamma-i\Omega)
(e^{i\Omega t}-e^{-\Gamma t})
\right]
=&
\Gamma\sin(\Omega t)-\Omega\cos(\Omega t)
+\Omega e^{-\Gamma t},
\end{align}
we yield for the inhomogeneous solution:
\begin{align}
\rho^I_{BB}(t)
=&
-iB_0 
e^{-\Gamma t}
\left[
 \Gamma_n\cos(\Omega t)
+\Omega_n\sin(\Omega t)
-\Gamma_n
e^{-\Gamma t}
\right] 
\\
&
-
D_0
e^{-\Gamma t}
\left[
 \Gamma_n\sin(\Omega t)
-\Omega_n\cos(\Omega t)
+\Omega_n e^{-\Gamma t}
\right] \\
&
+
\Sigma_0
e^{-\Gamma t}
\left[
\Omega_n
\sinh(\Gamma t)
-\Gamma_n
\sin(\Omega t)
\right]
. 
\end{align}
The complete solution is now given via:
\begin{align}
\rho_{BB}(t) 
=&
\rho_{BB}(0) 
e^{-2\Gamma t}
+
-iB_0 
\frac{\Omega}{2}
e^{-\Gamma t}
\left[
 \Gamma_n\cos(\Omega t)
+\Omega_n\sin(\Omega t)
-\Gamma_n
e^{-\Gamma t}
\right] 
\\
&
-
D_0
\frac{\Omega}{2}
e^{-\Gamma t}
\left[
 \Gamma_n\sin(\Omega t)
-\Omega_n\cos(\Omega t)
+\Omega_n e^{-\Gamma t}
\right] \\
&
+
\Sigma_0
\frac{\Omega}{2}
e^{-\Gamma t}
\left[
\Omega_n
\sinh(\Gamma t)
-\Gamma_n
\sin(\Omega t)
\right].
\end{align}
With the biexciton dynamics given, all other excitonic dynamics
can be directly calculated, using the relation given by the 
symmetries of the full set of equation of motion in Eq.~\eqref{eq:full_meq}.

\section{Convenient relations of the dressed state normalization 
factors}\label{app:symplectic_factors}
We define following normalization factors:
\begin{align}
a_1 = \frac{\Omega}{\sqrt{2\Omega^2+e^2_3}}, \qquad
a_2 = \frac{\Omega}{\sqrt{2\Omega^2+e^2_4}}, \qquad\text{and}\qquad
a_3 = \frac{1}{\sqrt{2}}. 
\end{align}
with
\begin{align}
e_3 =& \frac{1}{2} \left( \Delta + \sqrt{8\Omega^2+\Delta^2} \right), \qquad
e_4 = \frac{1}{2} \left( \Delta - \sqrt{8\Omega^2+\Delta^2} \right).
\end{align}
For example:
\begin{align}
e_3e_4=& 
\frac{1}{4}\left( \Delta^2 - 8\Omega^2 - \Delta^2 \right)
= -2\Omega^2 .
\end{align}
And,
\begin{align}
a_1^2+a_2^2 =& \frac{1}{2} \\
a^2_1 \Delta_+ + a^2_2 \Delta_- &= 0 \\
a^2_1 \Delta_- + a^2_2 \Delta_+ &= \frac{\Delta}{2} .
\end{align}
Just for convenience some more interesting algebraic relations:
\begin{align}
\frac{1}{2} \frac{e_3}{e_3 - e_4} =& a_2^2\\
\frac{1}{2} \frac{e_4}{e_4 - e_3} =& a_1^2 .
\end{align}
With these algebraic relations, orthogonality and orthonormalization
can easily be shown.

\section{Dressed correlation calculation} \label{app:dressed_correlation}
Here, we calculate the dynamics of the $+$-state photon, which 
depends on the following population dynamics:
\begin{align}
\rho^{++}_{++}(\tau) :=
\ew{\sig{+}{+}(\tau)}|_{\rho^{++}(0)} 
=&
\frac{1}{4}
\left(
\rho^{++}_{BB}(\tau)
+
\rho^{++}_{GG}(\tau) 
+
2
\text{Re}
\left[
\rho^{++}_{GB}(\tau) 
\right]
\right).
\end{align}
We can safely use $\rho_{GG}^{++}(\tau)=\rho_{BB}^{++}(\tau)-D^{++}(\tau)$,
then
\begin{align} 
4\rho^{++}_{++}(\tau) 
=&
2\rho^{++}_{BB}(\tau)
-
D^{++}(\tau) 
+
2
\text{Re}
\left[
\rho^{++}_{GB}(\tau) 
\right]
\\ \notag
=&
\phantom{+}
2
\left( 
\rho^{++}_{BB}(0)e^{-2\Gamma\tau}
-B_0^{++}\frac{-\Omega}{2}\left[...\right]
-D_0^{++}\frac{-\Omega}{2}e^{-\Gamma\tau}\left[...\right]
+\Sigma_0^{++}\frac{-\Omega}{2}e^{-\Gamma\tau}\left[...\right] 
\right) \\ \notag 
&
-\left(D_0^{++}+\Sigma_0^{++}\Gamma\Gamma_n\right) 
e^{-\Gamma\tau}\cos(\Omega\tau)
+\left(iB_0^{++}+\Sigma_0^{++}\Gamma\Omega_n\right)
e^{-\Gamma\tau} \sin(\Omega\tau)
+\Sigma_0^{++}\Gamma\Gamma_n 
\\ \notag 
& 
+ 2\text{Re}
\left[
\frac{B(\tau)}{2}
+
\frac{\rho^{++}_{BG}(0)+\rho^{++}_{GB}(0)}{2}
e^{-\Gamma\tau}
\right].
\end{align}
Now, we use the initial conditions given in the text:
$ D^{++}_0=0$, $ B^{++}_0=0$, $ \Sigma^{++}_0=2$ and 
$ \rho^{++}_{BB}(0)=1$, and $ \rho^{++}_{GB}(0)+\rho^{++}_{BG}(0)=2$.
Then, we see that $ B(\tau) $ is purely imaginary with these 
initial conditions.
The solution reads:
\begin{align} 
4\rho^{++}_{++}(\tau) 
=&
2 e^{-2\Gamma\tau}
\left[ 
1-\frac{1}{2}\Omega\Omega_n
\right] 
+2 e^{-\Gamma\tau}
\left[ 
1-\Gamma\Gamma_n \cos(\Omega\tau)
\right] 
+
\Omega\Omega_n
+2\Gamma\Gamma_n
.
\end{align}
The long-time limit is easily seen, and so can be the correlation
normalized and the $++$-detection probability derived.
The same calculation can be done for a cross correlation of the 
dressed states. 
For example, $ \rho^{00}_{++}(\tau) :=
\ew{\sig{+}{+}(\tau)}|_{\rho^{00}(0)}  $ differs to be calculation
above only in the inital conditions of 
$ \rho^{++}_{GB}(0)+\rho^{++}_{BG}(0)=-2$.
Therefore,
\begin{align} 
4\rho^{00}_{++}(\tau) 
=&
2 e^{-2\Gamma\tau}
\left[ 
1-\frac{1}{2}\Omega\Omega_n
\right] 
-2 e^{-\Gamma\tau}
\left[ 
1+\Gamma\Gamma_n \cos(\Omega\tau)
\right] 
+
\Omega\Omega_n
+2\Gamma\Gamma_n
.
\end{align}
with $ \rho^{00}_{++}(0)=0 $.
To calculate the experimental signal, we need furthermore 
the correlation functions, where the $0-$state photon is 
measured.
\begin{align}
\rho^{++}_{00}(\tau) :=
\ew{\sig{0}{0}(\tau)}|_{\rho^{++}(0)} 
=&
\frac{1}{4}
\left(
\rho^{++}_{BB}(\tau)
+
\rho^{++}_{GG}(\tau) 
-
2
\text{Re}
\left[
\rho^{++}_{GB}(\tau) 
\right]
\right).
\end{align}
It differs only in the sign of the transition dynamics.
However, due to the initial conditions it turns out that 
$ \rho^{++}_{00}(\tau) =\rho^{00}_{++}(\tau)$.
And, also $ \rho^{00}_{00}(\tau) =\rho^{++}_{++}(\tau)$,
since the different sign in the transition dynamics is 
changed due to the changed sign in the inital conditions.
Given all four combination of detection sequences, 
we can model the experimental signature, which is a superposition
of $ +,0-$detection events.
\end{appendix}

\noindent
\section*{References}

\bibliographystyle{unsrt}

\end{document}